# Biomarker of extracellular matrix remodelling C1M and proinflammatory cytokine IL-6 are related to synovitis and pain in end-stage knee osteoarthritis patients

Maja R Radojčić, MPharm, MSc
Nordic Bioscience Biomarkers & Research, Nordic Bioscience A/S, Herlev, Denmark
Department of Drug Design and Pharmacology, Faculty of Health and Medical Sciences, University of Copenhagen, Denmark
Christian S Thudium, PhD
Nordic Bioscience Biomarkers & Research, Nordic Bioscience A/S, Herlev, Denmark
Kim Henriksen, PhD
Nordic Bioscience Biomarkers & Research, Nordic Bioscience A/S, Herlev, Denmark
Keith Tan, PhD
Neuroscience, Innovative Medicines and Early Development, AstraZeneca, Cambridge, United Kingdom
Rolf Karlsten, MD, PhD
Department of Surgical Sciences, Anaesthesiology and Intensive Care, Uppsala University, Uppsala, Sweden
Amanda Dudley, PhD
Neuroscience, Innovative Medicines and Early Development, AstraZeneca, Cambridge, United Kingdom
Iain Chessell, PhD
Neuroscience, Innovative Medicines and Early Development, AstraZeneca, Cambridge, United Kingdom
Morten A Karsdal, PhD
Nordic Bioscience Biomarkers & Research, Nordic Bioscience A/S, Herlev, Denmark
Anne-Christine Bay-Jensen, PhD
Nordic Bioscience Biomarkers & Research, Nordic Bioscience A/S, Herlev, Denmark
Michel D Crema, MD
Quantitative Imaging Center, Department of Radiology, Boston University School of Medicine, Boston, Massachusetts, USA
Department of Radiology, Hôpital Saint-Antoine, University Paris VI, Paris, France
Ali Guermazi, MD, PhD
Quantitative Imaging Center, Department of Radiology, Boston University School of Medicine, Boston, Massachusetts, USA

Number of pages: 32; Tables: 5; Figures: 4.
*Correspondence: Maja R Radojčić – Nordic Bioscience Biomarkers & Research, Herlev Hovedgade 205-207, Herlev, 2730 Denmark; Tel: +45 4452 5252; E-mail: mra@nordicbio.com; URL: www.nordicbioscience.com

DISCLOSURE
Preliminary results of this study were included in the abstract that was presented in the form of a poster
presentation at the 16th World Congress on Pain, September 2016, Yokohama, Japan.

CONFLICT OF INTEREST
MRR is Marie Sklodowska-Curie fellow and has no competing financial interest in relation to the work described. KH, MAK, ACBJ, and CST are Nordic Bioscience employees; KH, MAK and ACBJ own stocks at Nordic Bioscience. KT, IC, and AD are AstraZeneca employees. RK was an AstraZeneca employee when the clinical phase of the study was conducted. AG is president of Boston Imaging Core Lab (BICL), a company
providing radiologic image assessment services, and consultant to MerckSerono, OrthoTrophix, AstraZeneca, Genzyme and TissueGene. MDC is a shareholder of BICL.

ABSTRACT

Little is known about local and systemic biomarkers in relation to synovitis and pain in end-stage osteoarthritis (OA) patients. We investigated the associations between the novel extracellular matrix biomarker, C1M, and local and systemic interleukin 6 (IL-6) with synovitis and pain. Serum C1M, plasma and synovial fluid IL-6 (p-IL-6, sf-IL-6) were measured in 104 end-stage knee OA patients. Contrast-enhanced magnetic resonance imaging (MRI) was used to semi-quantitatively assess an 11-point synovitis score; pain was assessed by the Western Ontario & McMaster Universities Osteoarthritis Index (WOMAC) and the Neuropathic Pain Questionnaire (NPQ). Linear regression was used to investigate associations between biomarkers and synovitis, and biomarkers and pain while controlling for age, sex and body mass index. We also tested whether associations between biomarkers and pain were confounded by synovitis. We found sf-IL-6 was associated with synovitis in the parapatellar subregion (B=0.006; 95% CI 0.003-0.010), and no association between p-IL-6 and synovitis. We also observed an association between C1M and synovitis in the peri-ligamentous subregion (B=0.013; 95% CI 0.003-0.023). Further, sf-IL-6, but not p-IL-6, was significantly associated with pain, WOMAC (B=0.022; 95% CI 0.004-0.040) and NPQ (B=0.043; 95% CI 0.005-0.082). There was no association between C1M and WOMAC pain but we did find an association between C1M and NPQ (B=0.229; 95% CI 0.036-0.422). Lastly, synovitis explained both biomarker-NPQ associations, but not the biomarker-WOMAC association. These results suggest C1M and IL-6 are associated with synovitis and pain, and synovitis is an important confounding variable when studying biomarkers and neuropathic features in OA patients.

# 1. INTRODUCTION

Osteoarthritis (OA) is a common disorder of synovial joints manifested by pain, stiffness, and general malfunction of the affected joints. Its pathophysiology is not clearly established, and pain is the most disabling symptom [3]. Pain is also the main reason why individuals seek medical care and a major factor in loss of joint function [3] leading to surgical intervention, e.g. total knee replacement (TKR).

All knee tissues are affected by OA and almost all are potential sources of pain, except the noninnervated articular cartilage. Typical changes in OA-affected joints include progressive degeneration ofthe articular cartilage, increased subchondral bone remodelling [14], synovial membrane inflammation [31], and meniscal and ligamentous injuries [29]. Magnetic resonance imaging (MRI) is a frequently used method for evaluating the contribution of synovitis to OA pathophysiology, and different tools are available for quantification of several important characteristics of synovitis [13; 16; 19]. Additionally, contrast-enhanced MRI provides more specific assessment of synovitis than non-contrast-enhanced MRI [15].Furthermore, the synovial membrane is innervated by unmyelinated C-fibres indicating its role as a potential pain generator [34]. C-fibres in joints exhibit sensitisation in inflammatory conditions and OA,with decreased thresholds and a stronger response to mechanical stimulation [32; 34]. This mechanical hyperalgesia is the main characteristic of nociceptive joint pain. There is, however, increasing evidence of a neuropathic component of OA pain [10; 11; 40], the type of pain caused by lesions of the somatosensory system [4]. The neuropathic-like phenotype in OA has been associated with signs of central sensitisation [17; 18] which may augment pain severity [2].

Biomarkers that can evaluate more subtle, OA-specific changes at a molecular level are needed. Biomarkers of the extracellular matrix turnover could help to understand the pathophysiological processes that cause pain and joint failure. C1M is a biomarker [27] that has been found to be released in an ex vivo model of synovitis [24]. Further, it has been correlated with inflammation in rheumatoid arthritis [30; 37; 39], ankylosing spondylitis [36] and OA [38]. However, inflammation also involves cytokines which are essential in cell-cell communication and important mediators in the aberrant metabolism of joint tissues in OA. The pro-inflammatory cytokine interleukin 6 (IL-6) has been found to play an important role in inflammation, arthritis and mechanical hypersensitivity [12; 21]. Significantly, hyperalgesia induced by IL-6 is persistent and difficult to reverse [6; 7; 33] indicating a potential role in chronic pain. However, there are still no reports of a relationship between C1M and pain and a possible difference between local and systemic IL-6 in predicting synovitis and pain in OA patients.

Thus, we aimed to explore associations between C1M and IL-6 with synovitis and pain, common OA pain and neuropathic features, in patients selected for TKR. Firstly, we hypothesised that the extracellular matrix biomarker C1M is associated with synovitis assessed by contrast-enhanced MRI, and with both types of pain. Secondly, we hypothesized that local and systemic levels of IL-6 are associated with synovitis and pain in these patients.

## 2. METHODS

### 2.1. Study sample

This exploratory biomarker study (AstraZeneca Study D2285M00029; Clinicaltrials.gov ID: NCT01611441) was designed to describe patients with knee OA undergoing TKR surgery using local and systemic biomarker levels and MRI. The overall aim of the study was to find a clinically feasible and minimally invasive tool for the potential characterisation of OA patients. The study was performed in accordance with the ethical principles of the Declaration of Helsinki and according to national and international regulations.

All participants provided their written informed consent.

This study had explicit inclusion and exclusion criteria. Inclusion criteria were: patient scheduled for TKR; man or woman ≥40 years of age; symptomatic OA of the knee for at least 6 months prior to enrolment according to clinical and radiological American College of Rheumatology criteria; Western Ontario & McMaster Universities Osteoarthritis Index (WOMAC) pain on walking ≥40 mm; calculated creatinine clearance >60 mL/min; and willingness and ability to discontinue systemic and topical nonsteroid anti-inflammatory analgesics for at least 5 half-lives (wash out period). Exclusion criteria included: a current diagnosis of another form of arthritis; arthroscopy performed on the target knee within 3 months of enrolment; fibromyalgia or wide spread chronic pain not related to OA; use of immunosuppressive medication including steroids or intra-articular hyaluronic acid injections within 3 months before enrollment; history of immunodeficiency; history and/or presence of any clinically significant disease or disorder (including hepatic, renal, cardiac, metabolic, gastrointestinal, pulmonary, haematological, neurological) which has not been stable over the previous 3 months, or which may put the patient at risk by participating in the study, influence the results or influence the patient's ability to fully participate; TKR in the contralateral knee; significant abnormalities on the clinical assessment of the target knee; use of erythropoietin, plasma/blood donation or blood loss >500 mL during the 3 months prior to enrollment; previous allogeneic bone marrow transplant; contraindications for MRI; allergy to the contrast agent; or participation in another clinical trial.

The study was conducted in Sweden and Canada at six sites, and was organised into three visits over 3–20 days depending on the washout period. The first visit included obtaining informed consent at enrollment, checking the inclusion and exclusion criteria, and defining the washout period. The second visit, one to five days prior to surgery, included clinical examination, MRI investigation, and pain assessment using the WOMAC and the Neuropathic Pain Questionnaire (NPQ). The day of surgery when blood, plasma, serum, urine, synovial fluid, synovial membrane and cartilage for biomarker evaluation were collected was defined as the third visit. Candidates for TKR who completed all three visits were included in the study, 104 patients in all.

### 2.2. Biomarker measures

We used two biomarkers as determinants: extracellular matrix biomarker C1M and pro-inflammatory cytokine IL-6. C1M is the neo-epitope on position 755-764 of the $\alpha_1$-chain of the type I collagen cleaved by matrix metalloproteinases 2, 9 and 13 [27]. Type I collagen is the most abundant collagen and its turnover is a part of the normal maintenance of the extracellular matrix [22]. In inflammatory conditions, the net loss of type I collagen is elevated due to an increase in levels of matrix metalloproteinases. Therefore, C1M reflects inflammatory processes in connective tissues rich in type I collagen (bone, tendon, ligament, skin, sclera, cornea and blood vessels). Since C1M is destroyed by Cathepsin K, the osteoclast's main protease, it cannot be released from bone; thus, it is a soft connective tissue degradation biomarker. C1M was measured in serum samples using competitive enzyme-linked immunosorbent assay (ELISA) at Nordic Bioscience, Herlev, Denmark. Further, we measured local levels of IL-6 in synovial fluid (sf-IL-6), and systemic levels in citrate plasma samples (p-IL-6) [9] using immunoassays with electrochemiluminescence detection (Meso Scale Discovery) at AstraZeneca, Stockholm, Sweden. For statistical analysis, we first inspected for outliers, values outside the range $[Q_1 - 1.5 \times (Q_3 - Q_1), Q_3 + 1.5 \times (Q_3 - Q_1)]$, where Q1 and Q3 are the first and third quartile, respectively. We found no outliers among the C1M measures. However, we detected five outliers in the p-IL-6 measures (>6.80 pg/mL), and seven outliers in the sf-IL-6 measures (>594.00 pg/mL) (Supplemental Figure 1). The outliers were excluded from the analysis. They could be the result of measurement errors or other medical conditions which we were unable to control for, and which could influence the results and the conclusions.

Biomarkers had positively skewed distributions, meaning that the majority of patients had low levels; we used this as a guide for dichotomization. We created a low group - the first tertile of biomarker levels (C1M ≤25.23 ng/mL; p-IL-6 ≤1.37 pg/mL; sf-IL-6 ≤30.83 pg/mL), and a high group – the remaining two tertiles of biomarker levels (C1M >25.23 ng/mL; p-IL-6 >1.37 pg/mL; sf-IL-6 >30.83 pg/mL). Therefore, for statistical purposes, we used both, continuous and dichotomous measures.

2.3. MRI assessment

MRI was obtained with a standard knee coil with the subject supine using 1.5T scanners (Signa Excite HD 1.5T, GE Healthcare, Milwaukee, Wisconsin; Achieva 1.5T, Philips Medical Systems, Best, the Netherlands; or Aera 1.5T, Siemens, Erlangen, Germany). First, intermediate-weighted fast spin-echo images with fat suppression (IwFS) in all three planes were acquired (TR=3,600ms, TE=30ms, slice thickness=3mm, matrix=512x512, FOV=16). Contrast-enhanced MRI was performed after non-enhanced MRI. Renal function was checked for the safety of the subjects. Venous blood samples were collected during the first visit and serum creatinine was measured at a central laboratory vendor, Quintiles (Quintiles Laboratories Europe, The Alba Campus, Rosebank, Livingston, United Kingdom). Creatinine clearance was calculated according to the Cockroft-Gault method: calculated creatinine clearance (mL/min)=□140-age(years)□× Weight (kg) 72 ×serum creatinine (mg/dL) [×0.85 (for women)]

Only patients whose calculated creatinine clearance was above 60 mL/min were selected for pre- and postcontrast MRI scanning. Intravenous gadolinium was administered at a dose of 0.2 mL/kg (0.1 mmol/kg) body weight and followed by a saline flush with 5 mL of saline solution. Injection took 30 seconds and imaging began 2 minutes later.Sagittal and axial T1-weighted fast spin echo images with fat suppression were acquired (TR=385ms, TE=12ms, slice thickness=3mm, matrix=384x384, FOV=16). Non-enhanced and contrast-enhanced sequences were anonymised and centrally read by two experienced musculoskeletal radiologists (AG, MDC). The MRI osteoarthritis knee score (MOAKS) was used to semi-quantitatively score intercondylar Hoffa-synovitis and whole knee effusion-synovitis [19]. Hoffasynovitis, on sagittal IwFS images, presents ill-defined high-signal intensity of Hoffa's fat pad at the intercondylar region on fluid-sensitive sequences [8]. Effusion-synovitis, on axial IwFS images, shows a combination of joint effusion and synovial thickening on fluid-sensitive sequences depending on the degree of joint cavity swollenness [8]. For volumetric evaluation of total synovial volume in mm$^3$ [13], the synovium was manually segmented using all sagittal T1wFS images, on a slice-by-slice basis by one reader (AFH, non-author) under the supervision and in consensus with an experienced musculoskeletalradiologist (AG). Segmentation was done using the 3D software ITK-snap (Penn Imaging Computing and Science Laboratory) [8]. Lastly, synovial thickness, measured in millimetres on contrast-enhanced MRI (axial and sagittal T1wFS images), was evaluated semi-quantitatively using a scoring system described by Guermazi et al [16] at 11 anatomical regions with subscores (parapatellar, peri-ligamentous and perimeniscal). We used continuous measures (11-point synovitis sum score 0–22; parapatellar subscore 0–10; peri-ligamentous subscore 0–4; peri-meniscal subscore 0–4) and a dichotomized 11-point synovitis scores. Since no one patient had a low synovitis score, our groups were: mild/moderate (11-point synovitis sum score 5–12; 34.6% patients) and severe (11-point synovitis sum score > 12; 65.4% patients) synovitis [16].

2.4. Pain assessment

We used two different questionnaires to assess osteoarthritic pain: WOMAC 48 hours recall, and the NPQ. The WOMAC consists of three subscales: pain, stiffness, and functional limitations; only the pain scale was used in this study. We used a visual analogue scale version of the WOMAC (VA3.1) [5], meaning that five WOMAC questions were scored on a 100 mm scale, starting from 0 mm (no pain at all) to 100 mm (extreme pain). The patients filled in the WOMAC on two occasions: at the first visit when a response of ≥40 mm to 'walking on a flat surface' was used as inclusion criteria, and at the second visit before imaging. Similarly, on the same two occasions patients filled in the NPQ, a self-administered questionnaire with 12 items representing various aspects or qualities of pain [25]. This instrument also uses a 100-point scale, where 0 represents no pain and 100 is the worst pain imaginable. In the present study, for the purpose of statistical analysis, only the questionnaires from the second visit were used. Further, for statistical purposes, scales of both instruments were transformed by dividing the sum scores by 10; this resulted in a transformed WOMAC pain score of 0–50 and a transformed NPQ score of 0-120. Also, we computed dichotomous pain variables. We dichotomized WOMAC pain similarly to our first outcome, the 11-point synovitis score, creating the mild/moderate group - first two tertiles of the WOMAC pain (transformed score ≤32.60 mm), and the severe group - the last tertile of the continuous measure (transformed score >32.60 mm). The NPQ is a screening tool that aids in differentiating neuropathic from non-neuropathic pain, and provides a calculation of the discriminant score for two groups; we computed these two groups for our analyses as well (neuropathic pain group 46.1%, non-neuropathic pain group 53.9% of patients) [25]. Defined in this way, the neuropathic pain group is likely to have neuropathic features, although that does not necessarily mean a diagnosed condition that requires further examinations.

2.5. Statistical analysis

We described the study sample using measures for location and measures of spread for Gaussian and non-Gaussian distribution. Further, Spearman's correlation analysis was performed to show relationships between the variables of interest. To examine the association between continuous biomarker levels C1M and IL-6 with synovitis we used linear regression analysis. The analyses were performed step by step presenting the unadjusted results (Model 1), and results adjusted for age, gender, and body mass index (BMI) (Model 2). Similarly, to examine the association between biomarkers and pain we used the same models in the linear regression analyses. Additionally, to examine the possible confounding effect of synovitis in these biomarker-pain relationships, we fitted an additional model adjusted for the 11-point synovitis score (Model 3). Furthermore, we used dichotomous variables and performed logistic regression analysis. Since this is an additional analysis we reported only the adjusted results: Model 2 was used in the synovitis analyses, and Model 3 was used in the pain analyses. Interpretation of the findings was primarily based on the results from the main analyses; additional analyses were intended to complement the main analyses and to show the results of probability function in addition to a linear function. This was done because the biomarkers, predictors in all analyses, are commonly used as both, continuous and dichotomous measures. This approach should allow comparison with results from a wide range of similar studies. We did not aim to compare biomarkers and synovitis as predictors of pain; synovitis was used as an outcome in the first group analyses, and as a confounding variable in the second group analyses when the association biomarker-pain was investigated. We reported beta coefficients (B) in linear regression analyses, odds ratios (OR) in logistic regression analyses, their 95% confidence interval (CI) and p-value. Betas represent an increase or decrease in Synovitis score (0-22), WOMAC pain score (0-50) or NPQ score (0-120) per unit of increase of C1M (ng/mL) or IL-6 (pg/mL). OR denotes the odds of having severe synovitis, severe pain measured by WOMAC, or neuropathic pain according to NPQ when having a high level of C1M or IL-6 compared to the low level of these biomarkers. Type I error level was set at 5% ($p<0.050$). No imputation was done for missing values, so the number of participants included in each analysis varies. All statistical analyses were performed using Statistical Package for the Social Sciences (SPSS), 23.0 (IBM, Armonk, New York, USA).

## 3. RESULTS

### 3.1. Descriptive Statistics

The study sample characteristics are shown in Table 1. Patients were 67 years old on average, 61.5% were women and their mean BMI was around 30 kg/m$^2$. In this sample of TKR patients, all three biomarker levels were positively skewed, the median C1M level was 31.44 ng/mL (interquartile range (IQR)=21.86-45.93), the median p-IL-6 was 1.71 pg/mL (IQR=1.09-2.54), and median sf-IL-6 was 71.05 pg/mL (IQR=21.35-150.00). Further, we assessed two outcome groups, synovitis and pain. Among MRI measures of synovitis, the main measure was the 11-point synovitis score; it was normally distributed (Kolmogorov-Smirnov statistics (K-S)=0.086$_{df=104}$, p=0.055) with a mean value 14.27 (standard deviation (SD)=3.67). We also took into account the parapatellar, peri-ligamentous and peri-meniscal synovitis subscores of this full score (Table 1). In this study sample, synovial volume (K-S=0.072$_{df=104}$, p=0.200) was on average 56.01 mL (SD=17.14). The MOAKS measures of synovitis, intercondylar Hoffa-synovitis, and whole knee effusion synovitis are semi-quantitative scores with values 0 to 3. Whole knee effusion-synovitis was prevalent with 74.0% of patients having grades 2 or 3, while Hoffa-synovitis was less prevalent with 64.4% of patients having grades 0 or 1. The second group of outcomes were pain assessments; WOMAC pain (K-S=0.039$_{df=104}$, p=0.200) with mean 28.75 (SD=9.18) and NPQ (K-S=0.086$_{df=104}$, p=0.057) with mean 40.83 (SD=19.97). We performed Spearman's correlations between variables included in this study (Table 2). We found a statistically significant positive correlation between C1M and p-IL-6 ($r_s$=0.27, p=0.017), but not between C1M and sf-IL-6 ($r_s$=0.08, p=0.502). C1M also positively correlated with the 11-point synovitis score ($r_s$=0.28, p=0.006) and with both pain assessments, WOMAC ($r_s$=0.20, p=0.049) and NPQ ($r_s$=0.22, p=0.027). Interestingly, we found that p-IL-6 positively correlated with biomarkers levels, with C1M and synovial fluid IL-6 ($r_s$=0.28, p=0.022), but not with any synovitis or pain assessment. Additionally, sf-IL-6 correlated positively with the 11-point synovitis score ($r_s$=0.28, p=0.012) and synovial volume ($r_s$=0.39, p<0.001), but not with the pain assessments. As expected, synovitis measures correlated positively and significantly with each other. Of these synovitis measures only the 11-point synovitis score correlated positively with pain measures, WOMAC ($r_s$=0.19, p=0.048) and NPQ ($r_s$=0.25, p=0.011). Furthermore, in this sample, we found that pain assessments WOMAC and NPQ were positively and strongly correlated ($r_s$=0.62, p<0.001).

### 3.2. Association between Biomarkers and Synovitis

We tested the association between C1M serum levels and the 11-point synovitis score using linear regression analysis and found a trend toward significance (Model 2: B=0.033; 95% CI -0.006, 0.072; p=0.096). We tested C1M in relation to synovitis subscores to check for possible subscore-driven effects. In this case, we found that C1M was positively associated with synovitis in the peri-ligamentous subregion (Figure 1) when confounding variables were taken into account (B=0.012; 95% CI 0.002, 0.022; p=0.018) (Table 3).

On the other hand, we did not find any association between p-IL-6 measures and synovitis measures, (Table 3). However, we did find an association between sf-IL-6 levels and the 11-point synovitis score (Model 2: B=0.009; 95% CI 0.002, 0.016; p=0.018), and subscore-specific analyses showed that this association was driven by synovitis in the parapatellar subregion (Figure 1) (adjusted B=0.006; 95% CI 0.002, 0.009; p=0.003) (Table 3). In addition, we found an association between sf-IL-6 and synovial volume in the unadjusted model (B=0.038; 95% CI 0.004, 0.072; p=0.029) and also after controlling for confounding variables (B=0.040; 95% CI 0.004, 0.076; p=0.030).

### 3.3. Association between Biomarkers and Pain

We examined the association between C1M serum levels and WOMAC pain using linear regression analysis and found a statistically significant association in the unadjusted model (B=0.101; 95% CI 0.006, 0.196; p=0.038). However, when confounding effects of age, gender and BMI were taken into account the association was not statistically significant (B=0.075; 95% CI -0.014, 0.164; p=0.096). Further, we also tested the association between C1M and NPQ and found a positive association in the adjusted model (B=0.229; 95% CI 0.036, 0.422; p=0.020). When we performed an additional Model 3, we found that the synovitis sum score explained the observed association (B=0.190; 95% CI -0.001, 0.380; p=0.051) (Table 4). Further, we studied the association between IL-6 and pain. We found no association between p-IL-6 and WOMAC pain, but there was a weak association between p-IL-6 and NPQ which was explained after controlling for the confounding effects of age, sex, and BMI (B=4.029, 95% CI −0.318, 8.376; p=0.069). In contrast, the association between sf-IL-6 and WOMAC pain was not statistically significant until confounding variables were taken into account (B=0.022, 95% CI 0.004, 0.040; p=0.016). This association between sf-IL-6 and WOMAC pain remained statistically significant even after correcting for the synovitis score (B=0.020, 95% CI 0.001, 0.038; p=0.038) (Table4). Similarly, the association between sf-IL-6 and NPQ was statistically significant only when confounding variables were taken into account (B=0.043, 95% CI 0.005, 0.082; p=0.026), but this association was explained by the synovitis score (B=0.031, 95% CI −0.007, 0.069; p=0.112).

### 3.4. Additional Analyses

To complement the main analyses, we used logistic regression to test the association between dichotomous biomarkers and dichotomous synovitis level. We found that patients with a high C1M level and a high sf-IL-6 level had the increased odds (OR=3.60; 95% CI 1.37, 9.46; p=0.009; and OR=3.88; 95% CI 1.30, 11.60; p=0.015, respectively) of having severe synovitis when compared to the referent low levels. We found no association between dichotomous measures of p-IL-6 and synovitis (Table 5). We used the same dichotomous approach for studying pain. We found no difference in WOMACmeasured pain between high and low levels of any of the biomarkers. However, we found that TKR candidates with high C1M levels had the increased odds of having neuropathic pain according to the NPQ (OR=3.52; 95% CI 1.26, 9.80; p=0.016) even when synovitis was included as a confounding variable. Thus, the discrepancy between the main and the additional analysis found in the C1M-NPQ relationship favoured the additional analysis. We found no association between dichotomous IL-6, local or systemic, and NPQ (Table 5). Discriminative properties of dichotomous biomarkers in a stacked bar graph style with percent showing outcome values within the low or high biomarker level are shown in Supplemental Figure 2.

### 4. DISCUSSION

We found that, in OA patients scheduled for TKR, the extracellular matrix biomarker C1M was associated with synovitis in the peri-ligamentous subregion, and with neuropathic features. However, this association was no longer significant when synovitis was introduced as a confounding variable into the model of C1M and neuropathic features. Further, we found that local, but not systemic measures of IL-6 were associated with synovitis in the parapatellar subregion, and with pain as assessed by both WOMAC and NPQ. Similarly, as with C1M, the association between synovial fluid IL-6 and neuropathic features was explained by synovitis (Figure2).

This cross-sectional study suggests that end-stage OA is local, low-grade inflammation associated with neuropathic symptoms. Synovitis is the main inflammatory characteristic of OA and we found in this study sample that all patients had synovitis to some extent, with 65.4% having severe synovitis. The presence of inflammation in patients selected for TKR is in agreement with previous reports describing end-stage OA patients [41].

This study investigated the important difference between local and systemic IL-6 levels in predicting synovitis assessed by contrast-enhanced MRI, unlike previous research which has been mainly focused on IL-6 in relation to radiographic OA. We only observed the relationship between local, synovial fluid IL-6 levels and synovitis. Interestingly, local IL-6 was additionally associated with synovial fluid volume that determines joint swelling confirming the importance of this cytokine in joint inflammation. Conversely, no association between plasma IL-6 and synovitis was observed. Findings on the contribution of IL-6 to OA pathophysiology are mixed: some studies showed a significant predictive function [28], while others reported no association between serum IL-6 and radiographic OA, and a negative relationship between local IL-6 levels and OA progression [20]. Other authors also reported no findings between systemic IL-6 levels and histological synovitis score in end-stage OA patients [41]. Systemic concentrations of IL-6 are measured in serum or plasma, yet IL-6 is the most stable in citrate plasma samples [9]. However, serum and plasma concentrations usually correlate highly, so the findings for systemic IL-6 measures should be comparable and provide similar conclusions. The lack of findings between plasma IL-6 measures and synovitis could be due to the altered properties of the synovium affecting the permeability of IL-6 between the local and the systemic compartments or more likely, contribution from other tissues to the systemic level.

Furthermore, we investigated the extracellular matrix biomarker C1M in relation to synovitis. The synovial membrane is a source of matrix metalloproteinases which generate this biomarker [31] and C1M has been found to be released from an ex vivo model of inflamed synovium [24]. These synovial biopsies are isolated from patients undergoing TKR, and this clinical study evaluates biomarkers and MRI measures in a comparable group of patients allowing comparison to different study reports and findings that confirm each other. Interestingly, a systemic measure of C1M provides a comparable amount of information as a local IL-6 measure when investigating synovitis. However, C1M and IL-6 were related to synovitis at different anatomical regions. The 11-point synovitis instrument scores inflammation in those parts of the synovium adjacent to bone (patella), soft connective tissue (ligament) and a fibrocartilaginous structure (meniscus). It has been found that IL-6 plays an important role in bone metabolism [26] and C1M is related to soft connective tissue, so our findings were expected. However, the peri-ligamentous subregion accounts for a small part of the sum score providing a possible explanation for why only a trend toward significance between C1M and the synovitis sum score was found. Lack of findings in the peri-meniscal subregion offers confirmation since none of the biomarkers would be expected to reflect the fibrocartilaginous compartment.

The most pronounced symptom and the important criterion for TKR is severe pain. To better describe and understand pain in end-stage OA, we investigated two different dimensions of pain, common WOMAC pain and a neuropathic component. WOMAC is the most widely used pain instrument in the OA field and defines pain while walking, resting and standing; we also used NPQ, a validated instrument [25] for assessing neuropathic symptoms.

Here, we report that extracellular matrix degradation is associated with neuropathic features in OA, and that there is a difference between synovial fluid and plasma IL-6 levels and their associations with these two components of OA pain. Measurements in synovial fluid, expected to reflect local joint events, were significantly associated with both pain dimensions, while systemic IL-6 levels were not associated with pain. Still, systemic C1M can substitute for the more invasive measurement of local IL-6 in describing neuropathic features. Showing that synovitis is a confounding variable in the biomarker-NPQ associations indicates that levels of C1M in serum and IL-6 in synovial fluid that predict pain depend on synovitis. To our knowledge, these three findings are novel in the OA field. Importantly, local IL-6 is associated with WOMAC pain independently of synovitis. This could mean that IL-6 describes OA changes in other knee compartments associated with walking or standing captured by the WOMAC pain scale. The lack of association between C1M and WOMAC pain further favours the possibility that inflammation and soft connective tissue degradation are more specifically associated with neuropathic symptoms. It has been reported that in early OA but not in advanced OA [35] serum IL-6 is associated with pain as assessed by the Japanese Knee Osteoarthritis Measure, which is comparable to WOMAC [1]. Our study adds that there is no association in end-stage OA. We can speculate that if inflammation is a driver of OA, then in the early stage IL-6 concentrations peak in synovial fluid and consequently the contribution of joint IL-6 to the total circulation level is more significant; thus, the systemic IL-6 measure could be used as a determinant of pain.

As the disease becomes chronic, IL-6 levels may decrease in synovial fluid thereby affecting systemic IL-6 levels; the systemic level accounts for IL-6 from other tissues and it describes joint pain to a smaller extent, often not significantly. This also suggests that in prospective OA studies, a quadratic, rather than a linear relationship between systemic IL-6 and pain could be investigated.

This study was designed with a number of important strengths. It used a homogeneous sample, allowed for a wash-out period to eliminate medication effects, it measured both local and systemic IL-6 levels, sought to investigate a novel biomarker C1M, assessed synovitis using contrast-enhanced MRI and evaluated two dimensions of pain. It does, of course, have limitations. First, this is a cross-sectional study; we described the study sample at a single time point and it provides no insight into associations regarding progression or prognosis. As a result, our conclusions should be restricted to describing end-stage OA patients. Second, having a homogeneous sample reduces inter-subject variability, which is an advantage in many ways, but a drawback in terms of external validity. Third, our sample size, 104, is rather moderate and small effects may not be detectable in a group of this size. Fourth, pressure pain thresholds could complement pain assessment. Further, OA in other joints was not assessed and we could not control for this in the analyses that included the systemic biomarker assessments. Lastly, we accounted for the influence of the most common confounding variables in the statistical analyses, but this study like other observational studies cannot exclude residual confounding which could be related to both, determinant and outcome.

In spite of some inevitable shortcomings, our study provides interesting findings that should be further investigated in well-designed studies. Although synovitis is an important descriptor of end-stage OA, what triggers the onset and drives disease progression remains to be determined. Further research is needed to understand the relationship between synovitis and neuropathic features in OA. It might be expected that subjects scheduled for TKR would provide a homogenous sample to study, however, it is more likely that different phenotypes drive the course of the disease, and defining these could improve targeted pharmacological research and eventually treatment. Biomarkers which reflect extracellular matrix remodelling have a potential to define different OA phenotypes, and could provide key insights to guide research on identifying mechanisms suitable for pharmacological intervention [23].

In conclusion, our study suggests that C1M and local IL-6, but not systemic IL-6 levels, are associated with synovitis and pain and that these biomarkers are not independently associated with neuropathic features when synovitis is taken into account. This suggests an important confounding effect of synovitis in the relationships of biomarkers and neuropathic features in OA patients.

ACKNOWLEDGEMENTS

We would like to thank the patients that agreed to participate in this study and the staff from the participating centres in Canada and Sweden. We thank Dr Amber Fotinos-Hoyer for her great contribution in quantitatively evaluating the synovial volume in all knees included in this study. The present study has received funding from the European Union's Horizon 2020 research and innovation programme under the Marie Sklodowska-Curie grant agreement No 642720. The C1M measurements were partly supported by the D-BOARD consortium funded by European Commission Framework 7 programme (EU FP7; HEALTH.2012.2.4.5-2, project number 305815, Novel Diagnostics and Biomarkers for Early Identification of Chronic Inflammatory Joint Diseases). The funding agency had no role in the design and conduct of the study, collection, management, analyses and interpretation of the data. The clinical study was carried out by AstraZeneca (Clinicaltrials.gov ID: NCT01611441). MRR is Marie Sklodowska-Curie fellow and has no competing financial interest in relation to the work described. CST, KH, MAK, and ACBJ are Nordic Bioscience employees; KH, MAK and ACBJ own stocks at Nordic Bioscience. AD, IC, and KT are AstraZeneca employees. RK was an AstraZeneca employee when the clinical phase of the study was conducted. AG is president of Boston Imaging Core Lab (BICL), a company providing radiologic image assessment services, and consultant to MerckSerono, OrthoTrophix, AstraZeneca, Genzyme and TissueGene. MDC is a shareholder of BICL.

Table 1 Descriptive statistics of the study sample

| Variable | Mean (SD) | Median (Interquartile range) | Min-Max values |
| --- | --- | --- | --- |
| Age | 66.78 (7.24) | 66.50 (61.25-73.00) | 50.00-82.00 |
| Sex, Women | 61.5% | | |
| BMI (kg/m$^2$) | 30.40 (5.21) | 29.29 (26.51-32.77) | 20.52-51.03 |
| C1M (ng/mL) | 36.81 (18.59) | 31.44 (21.86-45.93) | 20.00-116.00 |
| p-IL-6 (pg/mL) | 1.96 (1.09) | 1.71 (1.09-2.54) | 0.64-5.13 |
| sf-IL-6 (pg/mL) | 107.06 (109.85) | 71.05 (21.35-150.00) | 1.75-461.00 |
| 11-point synovitis score (0-22) | 14.27 (3.67) | 14.00 (12.00-17.00) | 5.00-22.00 |
| Parapattelar synovitis subscore (0-10) | 7.46 (1.92) | 8.00 (6.00-9.00) | 2.00-10.00 |
| Peri-ligamentous synovitis subscore (0-4) | 2.90 (0.94) | 3.00 (2.00-4.00) | 1.00-4.00 |
| Peri-meniscal synovitis subscore (0-4) | 2.75 (0.96) | 3.00 (2.00-3.75) | 0.00-4.00 |
| Synovial volume (mL) | 56.01 (17.14) | 53.61 (43.77-67.66) | 22.64-96.51 |
| MOAKS Intercondylar Hoffa-synovitis (0-3) | 1.31 (0.74) | 1.00 (1.00-2.00) | 0.00-3.00 |
| MOAKS Whole knee effusion-synovitis (0-3) | 1.91 (0.66) | 2.00 (1.00-2.00) | 1.00-3.00 |
| WOMAC Pain (0-50) | 28.75 (9.18) | 28.20 (21.85-35.70) | 6.80-49.90 |
| Neuropathic Pain (0-120) | 40.83 (19.97) | 38.00 (26.00-55.00) | 1.60-97.60 |

BMI – Body mass index; MOAKS – MRI Osteoarthritis Knee Score; WOMAC – Western Ontario & McMaster Universities Osteoarthritis Index.

Table 2 Spearman's correlation between variables of interest

| Variables | C1M | p-IL-6 | sf-IL-6 | 11-point Synovitis | Synovial Volume | Intercondylar Hoffa Synovitis | Whole Knee Effusion Synovitis | WOMAC | NPQ |
|---|---|---|---|---|---|---|---|---|---|
| **C1M** | | | | | | | | | |
| $r_s$ | 1.00 | **0.27** | 0.08 | **0.27** | 0.05 | 0.15 | 0.02 | **0.20** | **0.22** |
| p-value | . | 0.017 | 0.502 | 0.006 | 0.655 | 0.130 | 0.860 | 0.049 | 0.027 |
| N | 102 | 78 | 80 | 102 | 102 | 102 | 102 | 102 | 102 |
| **p-IL-6** | | | | | | | | | |
| $r_s$ | | 1.00 | **0.28** | 0.21 | 0.16 | 0.17 | 0.01 | 0.11 | 0.20 |
| p-value | | . | 0.022 | 0.059 | 0.169 | 0.135 | 0.957 | 0.333 | 0.080 |
| N | | 80 | 67 | 80 | 80 | 80 | 80 | 80 | 80 |
| **sf-IL-6** | | | | | | | | | |
| $r_s$ | | | 1.00 | **0.28** | **0.39** | 0.20 | 0.21 | 0.16 | 0.11 |
| p-value | | | . | 0.012 | <0.001 | 0.078 | 0.054 | 0.142 | 0.322 |
| N | | | 82 | 82 | 82 | 82 | 82 | 82 | 82 |
| **11-point Synovitis** | | | | | | | | | |
| $r_s$ | | | | 1.00 | **0.48** | **0.42** | **0.30** | **0.19** | **0.25** |
| p-value | | | | . | <0.001 | <0.001 | 0.002 | 0.048 | 0.011 |
| N | | | | 104 | 104 | 104 | 104 | 104 | 104 |
| **Synovial Volume** | | | | | | | | | |
| $r_s$ | | | | | 1.00 | **0.40** | **0.43** | 0.06 | 0.11 |
| p-value | | | | | . | <0.001 | <0.001 | 0.566 | 0.268 |
| N | | | | | 104 | 104 | 104 | 104 | 104 |
| **Intercondylar Hoffa Synovitis** | | | | | | | | | |
| $r_s$ | | | | | | 1.00 | **0.32** | 0.19 | 0.17 |
| p-value | | | | | | . | 0.001 | 0.059 | 0.090 |
| N | | | | | | 104 | 104 | 104 | 104 |
| **Whole Knee Effusion Synovitis** | | | | | | | | | |
| $r_s$ | | | | | | | 1.00 | 0.01 | -0.02 |
| p-value | | | | | | | . | 0.891 | 0.863 |
| N | | | | | | | 104 | 104 | 104 |
| **WOMAC** | | | | | | | | | |
| $r_s$ | | | | | | | | 1.00 | **0.62** |
| p-value | | | | | | | | . | <0.001 |
| N | | | | | | | | 104 | 104 |
| **NPQ** | | | | | | | | | |
| $r_s$ | | | | | | | | | 1.00 |
| p-value | | | | | | | | | . |
| N | | | | | | | | | 104 |

$r_s$ - Spearman's correlation coefficient; N – number of patients included in an analysis; WOMAC – Western Ontario & McMaster Universities Osteoarthritis Index; NPQ – Neuropathic Pain Questionnaire.

Table 3 Association between extracellular matrix biomarker C1M and pro-inflammatory cytokine IL-6 with Synovitis assessed by contrast enhanced-MRI

| Outcome | | C1M (serum) | | IL-6 (plasma) | | IL-6 (synovial fluid) | |
|---|---|---|---|---|---|---|---|
| | | B (95%CI) | p-value | B (95%CI) | p-value | B (95%CI) | p-value |
| 11-point Synovitis sum score | Model 1 | 0.038 (0.001, 0.077) | 0.051 | 0.500 (-0.259, 1.258) | 0.194 | **0.010** (0.003, 0.016) | 0.008 |
| | Model 2 | 0.033 (-0.006, 0.072) | 0.096 | 0.391 (-0.439, 1.221) | 0.351 | **0.009** (0.002, 0.016) | 0.018 |
| *Subscores* | | | | | | | |
| Synovitis in Parapatellar subregion | Model 1 | 0.017 (-0.004, 0.037) | 0.110 | 0.273 (-0.114, 0.661) | 0.164 | **0.006** (0.003, 0.010) | 0.001 |
| | Model 2 | 0.013 (-0.007, 0.033) | 0.197 | 0.172 (-0.244, 0.589) | 0.412 | **0.006** (0.002, 0.009) | 0.003 |
| Synovitis in Peri-ligamentous subregion | Model 1 | **0.013** (0.003, 0.023) | 0.010 | 0.103 (-0.086, 0.293) | 0.281 | 0.001 (-0.001, 0.003) | 0.347 |
| | Model 2 | **0.012** (0.002, 0.022) | 0.018 | 0.067 (-0.141, 0.275) | 0.524 | 0.001 (-0.002, 0.002) | 0.634 |
| Synovitis in Peri-meniscal subregion | Model 1 | 0.007 (-0.003, 0.018) | 0.147 | 0.058 (-0.127, 0.242) | 0.535 | 0.001 (-0.001, 0.003) | 0.246 |
| | Model 2 | 0.007 (-0.003, 0.018) | 0.168 | 0.076 (-0.128, 0.281) | 0.461 | 0.001 (-0.001, 0.003) | 0.219 |

In C1M analyses 102 patients were included; IL-6 in plasma was measured in 80 patients; in IL-6 analyses, using measures in the synovial fluid, 82 were included. Synovitis sum score is whole-knee synovitis assessed at 11 sites; this score is the sum of three different anatomical subregions namely, parapatellar, peri-ligamentous and peri-meniscal. Models were constructed using linear regression analysis. Betas represent increase or decrease in Synovitis score for every unit of increase of C1M or IL-6.
Model 1 presents unadjusted results.
Model 2 presents results adjusted for age, sex and the body mass index (BMI).

Table 4 Association between extracellular matrix biomarker C1M and pro-inflammatory cytokine IL-6 with Pain assessed by WOMAC and NPQ

| Outcome | | C1M (serum) | | IL-6 (plasma) | | IL-6 (synovial fluid) | |
|---|---|---|---|---|---|---|---|
| | | B (95%CI) | p-value | B (95%CI) | p-value | B (95%CI) | p-value |
| **WOMAC Pain** | Model 1 | **0.101** (0.006, 0.196) | 0.038 | 1.213 (-0.624, 3.050) | 0.192 | 0.016 (-0.002, 0.035) | 0.082 |
| | Model 2 | 0.075 (-0.014, 0.164) | 0.096 | 1.351 (-0.483, 3.186) | 0.146 | **0.022** (0.004, 0.040) | 0.016 |
| | Model 3 | 0.068 (-0.022, 0.158) | 0.137 | 1.194 (-0.633, 3.020) | 0.197 | **0.020** (0.001, 0.038) | 0.038 |
| **NPQ** | Model 1 | **0.278** (0.078, 0.478) | 0.007 | **4.464** (0.368, 8.559) | 0.033 | 0.030 (-0.009, 0.070) | 0.127 |
| | Model 2 | **0.229** (0.036, 0.422) | 0.020 | 4.029 (-0.318, 8.376) | 0.069 | **0.043** (0.005, 0.082) | 0.026 |
| | Model 3 | 0.190 (-0.001, 0.380) | 0.051 | 3.385 (-0.794, 7.564) | 0.111 | 0.031 (-0.007, 0.069) | 0.112 |

Western Ontario and McMaster Osteoarthritis Index (WOMAC) and Neuropathic Pain Questionnaire were used for pain assessment in patients selected for total knee replacement (TKR). In C1M analyses 102 patients were included; in IL-6 analyses including measures in plasma 80 patients were included, and IL-6 in the synovial fluid was measured in 82 patients. Models were constructed using linear regression analysis. Betas represent increase or decrease in Synovitis score for every unit of increase of C1M or IL-6.
Model 1 presents unadjusted results.
Model 2 is adjusted for age, sex and the body mass index (BMI).
Model 3 is additionally adjusted for the synovitis sum score.

Table 5 Results of logistic regression between extracellular matrix biomarker C1M and pro-inflammatory cytokine IL-6 with Synovitis assessed by contrast-enhanced MRI and Pain assessed by WOMAC and NPQ

| Determinant | Outcome | 11-point Synovitis [a] (mild/moderate vs severe synovitis) | | WOMAC [b] (mild/moderate vs severe pain) | | NPQ [b] (non-neuropathic vs neuropathic pain) | |
|---|---|---|---|---|---|---|---|
| | | OR (95% CI) | p-value | OR (95% CI) | p-value | OR (95% CI) | p-value |
| **C1M (serum)** | High | **3.60** (1.37, 9.46) | 0.009 | 1.99 (0.66, 6.01) | 0.221 | **3.52** (1.26, 9.80) | 0.016 |
| | Low | - | - | - | - | - | - |
| **IL-6 (plasma)** | High | 2.00 (0.66, 6.15) | 0.221 | 0.88 (0.25, 3.07) | 0.840 | 1.40 (0.42, 4.64) | 0.587 |
| | Low | - | - | - | - | - | - |
| **IL-6 (synovial fluid)** | High | **3.88** (1.30, 11.60) | 0.015 | 2.00 (0.62, 6.40) | 0.245 | 1.27 (0.43, 3.76) | 0.670 |
| | Low | - | - | - | - | - | - |

Models were constructed using logistic regression analysis. Odds ratio (OR) represents the odds of having severe Synovitis, or severe pain measured by WOMAC, or neuropathic pain when having a high level of C1M or IL-6 compared to the low level of these biomarkers.
[a] Model is adjusted for age, sex and the body mass index (BMI).
[b] Model is adjusted for age, sex, BMI and the 11-point synovitis sum score.

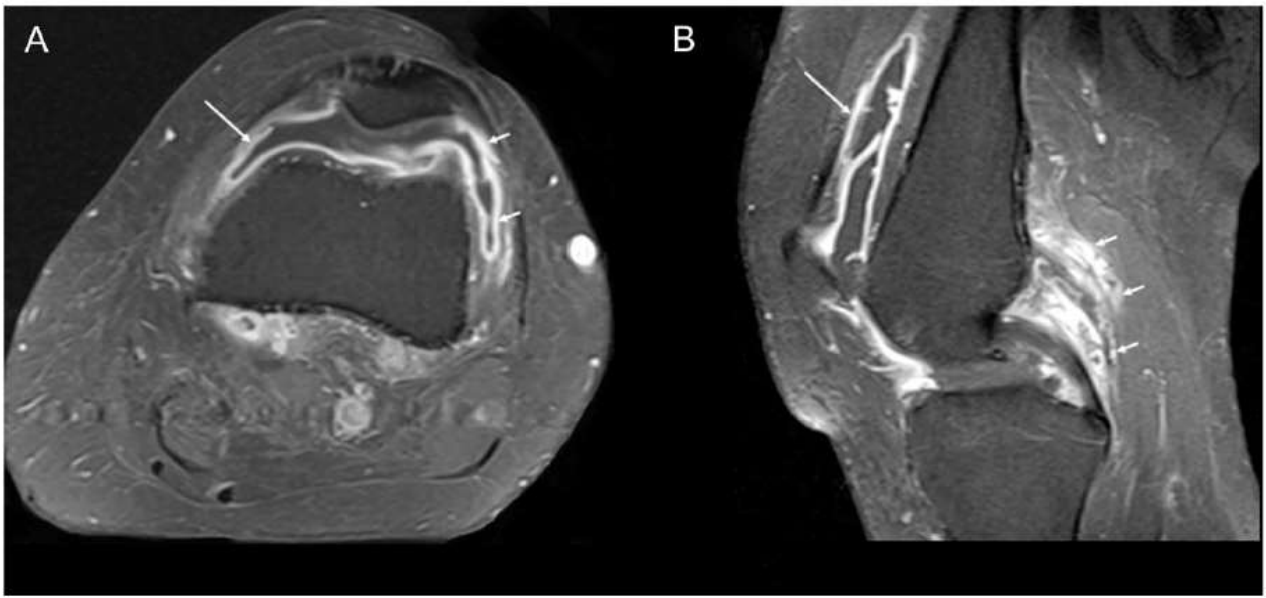

Figure 1 Synovitis seen on fat-suppressed contrast-enhanced MRI; a) axial view - mild medial (arrow) and moderate lateral (small arrows) parapatellar synovitis; b) sagittal view - moderate suprapatellar (arrow) and severe synovitis posterior to posterior cruciate ligament (small arrows).

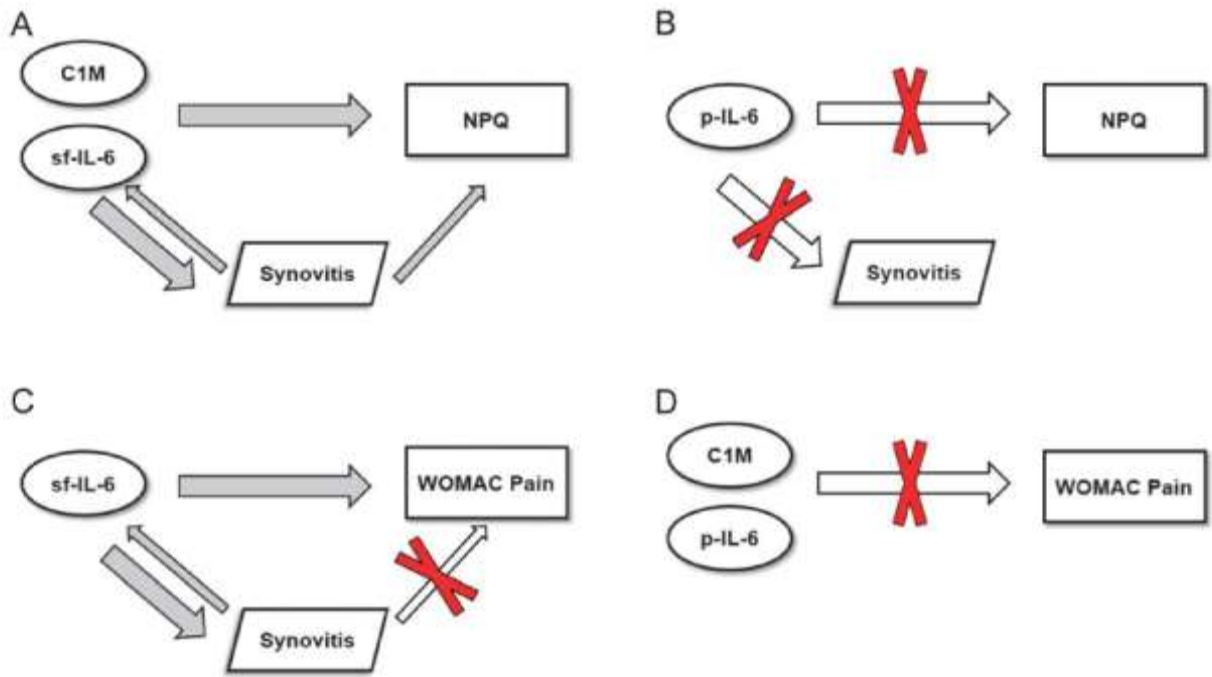

Figure 2 Schematic presentation of the findings from this study; a) serum C1M and synovial fluid IL-6 (sf-IL-6) are associated with synovitis and neuropathic features (NPQ), and synovitis is a confounding variable in the relationships biomarkers-NPQ; the latter tells that biomarker levels that predict NPQ depend on the synovitis score; b) plasma IL-6 (p-IL-6) is not associated with synovitis and NPQ; c) sf-IL-6 is associated with WOMAC pain, and synovitis is not a confounding variable in the relationship biomarker-WOMAC pain; the latter tells that biomarker predicts WOMAC pain independently of synovitis; d) C1M and p-IL-6 are not associated with WOMAC pain.

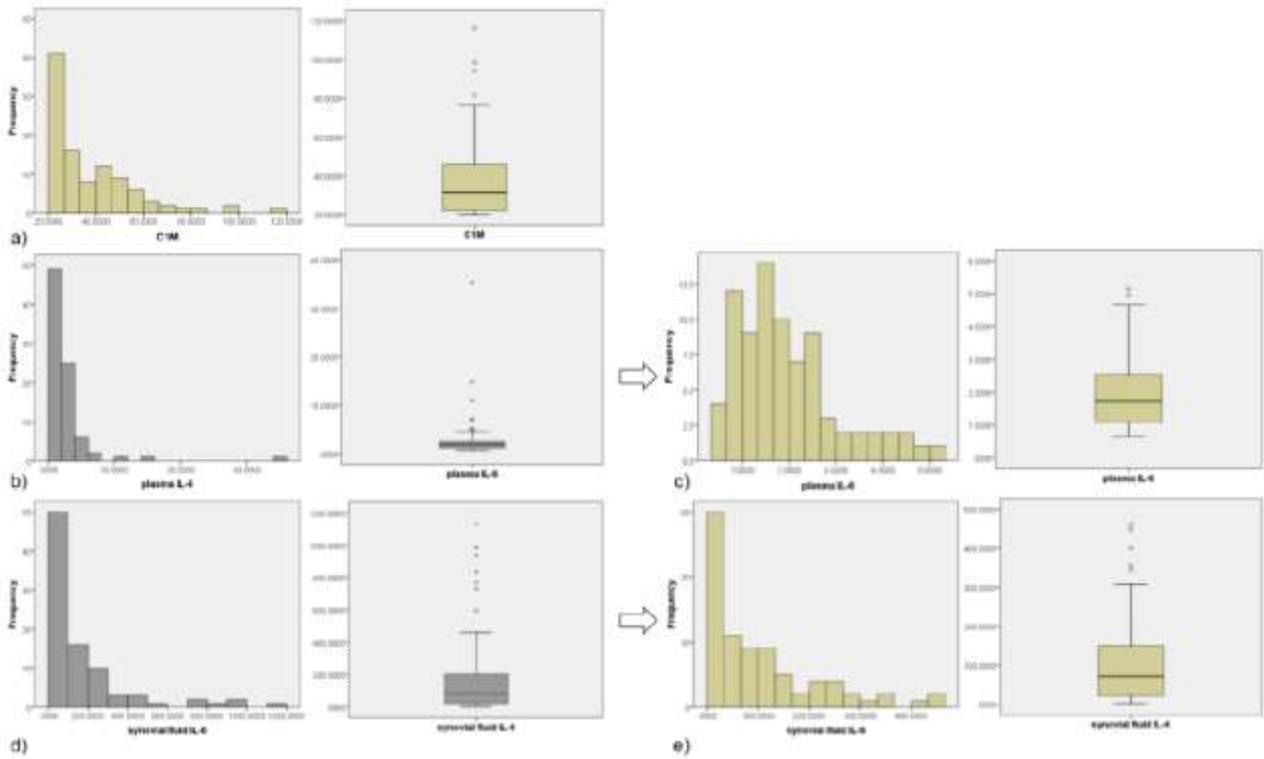

Supplemental Figure 1 Distributions of biomarkers are shown on histogram (left) and boxplot (right) graphs; stars indicate outliers, and circles indicate extreme values that are not outliers; a) C1M; b) plasma IL-6 - original data; c) plasma IL-6 - after excluding 4+1 outliers in two steps; d) synovial fluid IL-6 - original data; e) synovial fluid IL-6 - after excluding 5+2 outliers in two steps.

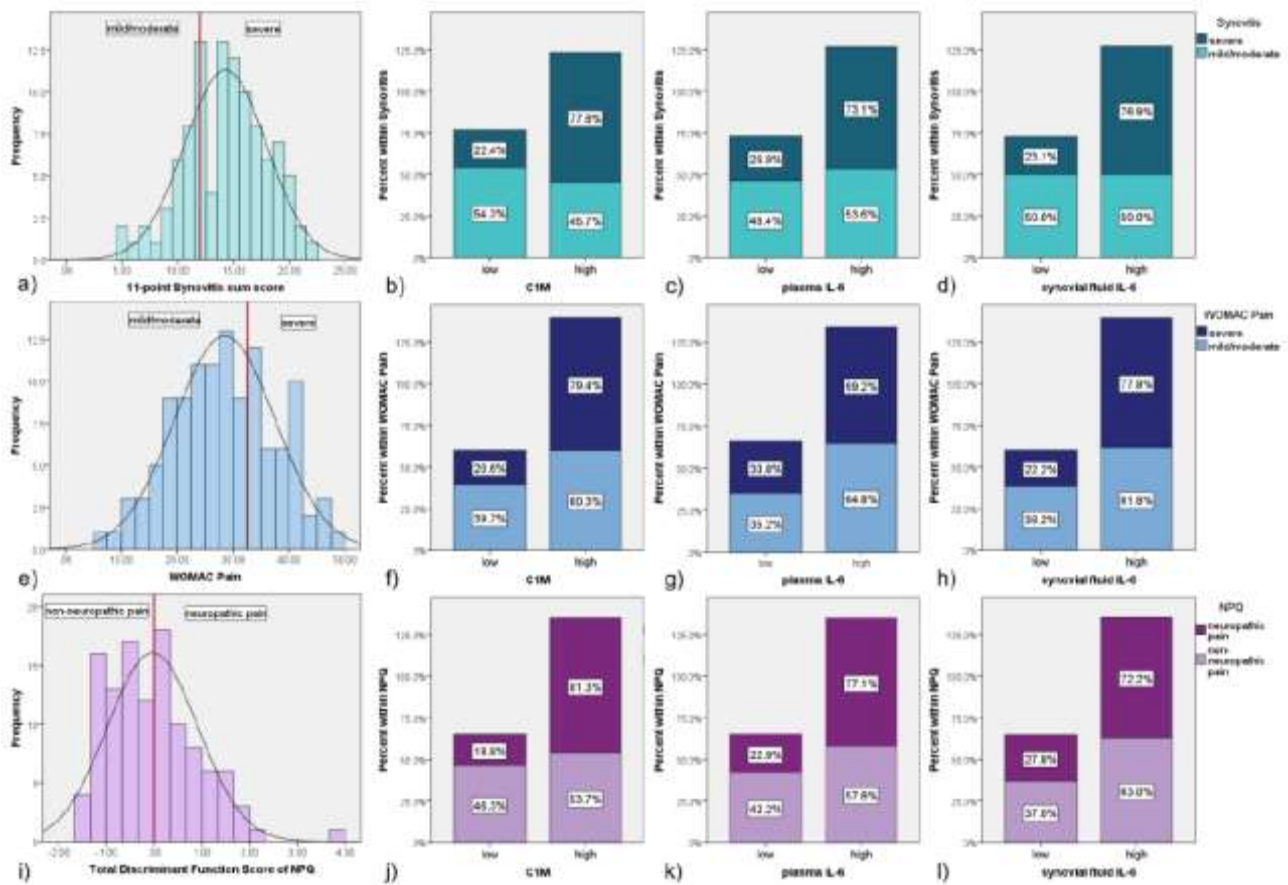

Supplemental Figure 2 Dichotomization of the outcome variables and discrimination between these values by dichotomous biomarkers. Low level is the first tertile of a biomarker distribution; high level is the second and third tertiles. Histogram shows distribution and the red line indicates the cut-off for dichotomization; in the stacked bar graphs, the stacks represent the percent of the outcome values within low or high biomarker level; A light colour indicates mild/moderate synovitis or WOMAC pain, or nonneuropathic pain according to Neuropathic Pain Questionnaire (NPQ), and a dark colour indicates severe synovitis or WOMAC pain, or neuropathic pain according to NPQ. Biomarker sensitivity is shown by a dark coloured stack in a high biomarker bar while biomarker specificity is indicated by a light coloured stack in a low biomarker bar; the larger the two stacks the better the sensitivity and specificity of the biomarker; a) distribution of the 11-point synovitis sum score and dichotomization to mild/moderate and severe synovitis; b-d) discrimination between mild/moderate and severe synovitis by low and high serum C1M, plasma IL-6 and synovial fluid IL-6 levels; e) distribution of the WOMAC pain score and dichotomization to mild/moderate and severe WOMAC pain; f-h) discrimination between mild/moderate and severe WOMAC pain by low and high serum C1M, plasma IL-6 and synovial fluid IL-6 levels; i) distribution of the total discriminant function score of the NPQ that is used for dichotomization to non-neuropathic and neuropathic pain according to the NPQ; note: this score is derived from the NPQ, this is not a distribution of the NPQ; j-l) discrimination between non-neuropathic and neuropathic pain according to NPQ by low and high serum C1M, plasma IL-6 and synovial fluid IL-6 levels. In all cases, serum C1M has the highest sensitivity and specificity compared to plasma and synovial fluid IL-6, yet all three biomarkers have quite low specificity.